# Quantum interference between light sources separated by 150 million kilometers

Yu-Hao Deng[1,2], Hui Wang[1,2], Xing Ding[1,2], Z.-C. Duan[1,2], Jian Qin[1,2], M.-C. Chen[1,2], Yu He[1,2], Yu-Ming He[1,2], Jin-Peng Li[1,2], Yu-Huai Li[1,2], Li-Chao Peng[1,2], E. S. Matekole[3], Tim Byrnes[4], C. Schneider[5], M. Kamp[5], Da-Wei Wang[6], Jonathan P. Dowling[1,3,4], Sven Höfling[1,5,7], Chao-Yang Lu[1,2], Marlan O. Scully[8,9,10], and Jian-Wei Pan[1,2]

[1]Shanghai Branch, Department of Modern Physics and National Laboratory for Physical Sciences at the Microscale, University of Science and Technology of China, Shanghai 201315, China
[2]CAS Center for Excellence and Synergetic Innovation Center in Quantum Information and Quantum Physics, University of Science and Technology of China, Shanghai 201315, China
[3]Hearne Institute for Theoretical Physics and Department of Physics and Astronomy, Louisiana State University, Baton Rouge, Louisiana 70803, USA
[4]New York University Shanghai, 1555 Century Ave, Pudong, Shanghai 200122, China
[5]Technische Physik, Physikalisches Instität and Wilhelm Conrad Röntgen-Center for Complex Material Systems, Universitat Würzburg, Am Hubland, D-97074 Würzburg, Germany
[6]Department of Physics, Zhejiang University, Hangzhou 310027, China
[7]SUPA, School of Physics and Astronomy, University of St. Andrews, St. Andrews KY16 9SS, United Kingdom
[8]Institute for Quantum Science and Engineering, Texas A&M University, College Station, Texas 77843, USA
[9]Department of Physics, Baylor University, Waco, Texas 76798, USA
[10]Department of Mechanical and Aerospace Engineering, Princeton University, Princeton, New Jersey 08544, USA
Corresponding authors: CYL cylu@ustc.edu.cn, JWP pan@ustc.edu.cn

**We report an experiment to test quantum interference, entanglement and nonlocality using two dissimilar photon sources, the Sun and a semiconductor quantum dot on the Earth, which are separated by ~150 million kilometers. By making the otherwise vastly distinct photons indistinguishable in all degrees of freedom, we observe time-resolved two-photon quantum interference with a raw visibility of 0.796(17), well above the 0.5 classical limit, providing the first evidence of quantum nature of thermal light. Further, using the photons with no common history, we demonstrate post-selected two-photon entanglement with a state fidelity of 0.826(24), and a violation of Bell's inequality by 2.20(6). The experiment can be further extended to a larger scale using photons from distant stars, and open a new route to quantum optics experiments at an astronomical scale.**

Can any two photons in the Universe, no matter how distantly and independently they originate from, show quantum interference and entanglement? According to quantum theory, when two quantum-mechanically indistinguishable single photons impinge upon a 50/50 beam splitter, they bunch together out of the same output port due to bosonic statistics. The classical picture of electromagnetic fields failed in understanding the interference of two photons from independent sources with a visibility better than 50%[1-4], which can be explained by quantum interference of the probability amplitudes of the two-photon events[5]. This effect, also known as Hong-Ou-Mandel (HOM) two-photon interference[6], poses a strong conceptual challenge to the celebrated statement by Dirac that "Each photon then interferes only with itself. Interference between different photons never occurs"[7].

Since the original HOM experiment, which used two photons from the same parametric down-converted pair, progressively more and more independent, dissimilar, and distant sources have been used to test this type of quantum interference[8-18]. Apart from its fundamental interest, the quantum interference of photons, emitted from different sources and without a common origin, can be harnessed for various quantum information tasks such as quantum teleportation[19], quantum repeaters[20], hybrid quantum networks[21], and measurement-device-independent quantum-key distribution[22].

Here, we demonstrate non-classical interference between sunlight, and single photons from a semiconductor quantum dot (See Fig. 1 for an illustration of the experimental setup). The sunlight is collected with the assistance of an electric-motor-driven equatorial mount (located at 121.5°E, 31.1°N , in Shanghai) that tracks the Sun. The collected sunlight is then guided into our laboratory through a 50-meter-long single-mode fiber. We choose days when there are few clouds and monitor the clouds' distribution and motions in real time to avoid intensity fluctuations.

We use a self-assembled InAs/GaAs quantum dot (QD) as an ultra-bright single-photon source with near-unity purity and indistinguishability[23,24]. The QD is embedded in a 2-μm diameter micropillar and sandwiched between 25.5 (15) λ/4-thick AlAs/GaAs mirror pairs that form lower (upper) distributed Bragg reflectors. At a temperature of 4 K, the QD is spectrally resonant with the mode of the microcavity yielding a Purcell factor of 7.6. The high Purcell factor mitigates dephasing and efficiently funnels the spontaneous emission of the QD into the cavity mode, which enhances the single-photon extraction efficiency to 82%. Under resonant excitation with a picosecond π-pulsed laser, with a repetition rate of 76 MHz, the source emits 25.6 million polarized single photons per second. After passing through an 1-GHz etalon, the photons at wavelength 893.198 nm (in air) show a second-order correlation function $g_Q^{(2)}(0) = 0.011(1)$ at zero-time delay (Fig. 2a). We perform a series of HOM experiments with two QD single photons with their emission time separation varying from 13 ns to 6.4 μs, and observe the photon's indistinguishability from 0.974(1) at 13 ns (see Fig. 2b) slightly dropping to a plateau of 0.952(3) through a dephasing time scale of ~100 ns (see Supplemental Material). We measure that the QD single photons are Fourier transform-limited with an indistinguishability of 0.952(3).

It is clear that the sunlight differs dramatically from the QD single-photon source in nearly all degrees of freedom, including polarization, spatial modes, spectral and temporal properties, and also photon statistics. To observe the HOM quantum interference, the two single photons from the independent sources should be made quantum mechanically indistinguishable. Firstly, the sunlight has an extremely broad and complex solar spectrum, which can become even more complicated after passing through Earth's atmosphere. We use optical filters and a temperature-tunable etalon with a bandwidth of 1 GHz to filter the sunlight so that it matches the spectrum of the QD single photons.

Secondly, for the temporal degree of freedom, while the QD single photons are pulsed and synchronized with the excitation picosecond laser, the sunlight is continuous and random in time. Thus, we have to rely on fast single-photon detection time resolution to discriminate and post-select the fraction of photon pairs that overlap on the beam splitter simultaneously. In this work, we use a superconducting nanowire single-photon detector with a time jitter of ~20 ps to register the single-photon events. The detection time resolution is much shorter than the photons' coherence time, dictated by the 1-GHz filtering bandwidth. The coincidence signals are then registered with a time-to-digital converter with a resolution of 1 ps. Moreover, to eliminate the unwanted multi-photon events from the sunlight (continuous in all the ~13 ns in each period) outside the time window of the QD actually emitting a single photon (~100 ps), we electrically gate the detections with a window of 1 ns, synchronized to the QD excitation laser pulses. This method suppresses the multi-photon contribution from the sunlight by a factor of ~13 compared to the case of ungated detection.

Thirdly, for a good spatial-mode overlap, we collect the single photons from both sources into single-mode optical fibers, which transforms the photons into the fundamental transverse Gaussian mode. For the polarization degree of freedom, the photons are prepared in a single polarized state using polarizing beam splitters before the two-photon interference.

Last but not least, there is a counterintuitive property of the light sources — which is not apparent in the first-order correlation function (but is manifest in their second-order correlation function)[25] — that can also significantly influence the two-photon interference visibility. Unlike the high-purity QD single-photon source, sunlight is intrinsically thermal light, theoretically predicted second-order correlation function of $g_\text{S}^{(2)}(0)=2$ (see Fig. 2c), indicating appreciable multi-photon-event contributions. In the case of an ideal (with zero multi-photon contribution) single-photon source interfering with a thermal light, if the intensity of the thermal light is sufficiently low, one would in principle obtain a near-unity interference visibility [16,26]. However, due to the small yet non-zero $g_\text{Q}^{(2)}(0)=0.011(1)$ of the QD single-photon source, the dependence of the raw visibility on the intensity of the thermal light becomes more complicated as shown in Fig. S5. We optimize the relative intensity of the two sources to achieve a maximal visibility (see Supplemental Material). In our experiment, the count rate of the photons from the Sun is set to be $2\times10^5$ s$^{-1}$, and from the QD is set as $1.5\times10^5$ s$^{-1}$.

Figure 3a shows the registered coincidence counts as a function of the time delay between the QD single photons and the sunlight photons for their relative polarizations prepared in orthogonal (red circle) and parallel (blue dot) directions. No background or dark-count subtraction is applied to the data. Each data point corresponds to a time bin of 10 ps. The red and blue curves are the result of our theoretical modeling, which takes into account the actual parameters of our experiment (see the Supplemental Material). For parallel polarizations, the data shows a pronounced dip at zero delay. We sum up coincidence counts that fall into a delay time-bin $\tau_{bin}$ between the two detectors to calculate the interference visibility. For $\tau_{bin}=20$ ps, a raw visibility of 0.796(17) is extracted, well above the classical limit of 0.5, which conclusively establishes the quantum nature of the two-photon interference between the QD and the Sun. The dependence of the visibility on the delay time-bin is illustrated in Fig. 3b, showing a visibility higher than 0.5 is maintained up to $\tau_{bin}=440$ ps. The non-vanishing dip at the zero-time delay is mainly attributed to the multi-photon events from the QD and the Sun, as we discuss further in the Supplemental Material.

While Hanbury Brown and Twiss interference with thermal light was demonstrated more than 60 years ago[27], their results could be explained within the framework of classical coherence theory. Previous attempts on HOM interference with table-top pseudo-thermal light sources have also been reported[16,28], but the low visibilities were insufficient to support the non-classical picture. In our experiment, however, the quantum nature manifests itself by directly observing visibilities well above 50%, which has no classical analog[1-4]. This is not a surprise for single-photon sources, since single-atom resonance fluorescence had shown anti-bunching 40 years ago that provided the first evidence of the quantum nature of light[25,29,30]. However, our result is the first time that a thermal light — requiring only classical optics for its description — is involved in a highly nonclassical quantum-optics experiment.

Owing to the combination of our high-performance single photon sources (developed only very recently), and techniques making the otherwise vastly different photons indistinguishable in all degrees of freedom, the observed raw visibility not only exceeds the classical limit of 0.5, for strictly proving non-classical interference, but also is higher than the threshold of 0.71 that enables post-selected entanglement generation and a test of Bell inequality. By initializing the two photons in orthogonal polarizations and overlapping them onto a beam splitter, and selecting only the events (with 50% probability) where there is one and only one photon in each output, the two output photons are projected into an entangled state close to the form of $|\psi^-\rangle = (|H\rangle_1|V\rangle_2 - |V\rangle_1|H\rangle_2)/\sqrt{2}$, where $H$ ($V$) denotes horizontal (vertical) polarization. We perform two-photon correlation measurements along the basis of $H/V$, $+/-$, and $R/L$, where $|\pm\rangle = (|H\rangle \pm |V\rangle)/\sqrt{2}$, and $|R\rangle = (|H\rangle + i|V\rangle)/\sqrt{2}$, $|L\rangle = (|H\rangle - i|V\rangle)/\sqrt{2}$. The data are plotted in Fig. 4, from which we calculate the state fidelity of the generated entanglement (defined as the wave-function overlap of the experimentally obtained state with the ideal $|\psi^-\rangle$) to be 0.826(24) for $\tau_{bin} = 50$ ps, which is sufficient to confirm the two-particle entanglement.

Finally, we use the two-photon entanglement thus generated to perform a Bell test in the Alley-Shih configuration[31]. While Bell experiments were usually performed with entangled particles sharing a common source, it is interesting to conduct Bell tests that use particles from physically separated remote sources sharing no common history[32]. Here, we use the QD single photons and the sunlight which were originally separated by one astronomical unit to test the Clauser-Horne-Shimony-Holt (CHSH)-type inequality[33], which is given by

$$S=|E(\varphi_1,\varphi_2)-E(\varphi_1,\varphi_2')+E(\varphi_1',\varphi_2)+E(\varphi_1',\varphi_2')|\leq 2,$$

where $E(\varphi_1,\varphi_2)$ is the two-photon correlation at measurement angles of $\varphi_1$ and $\varphi_2$ respectively. The angles are selected among (0, π/8), (0, 3π/8), (π/4, π/8) and (π/4, 3π/8). The observed data in the four settings are summarized in Fig. 5, from which we find $S=2.20\pm0.06$, with a violation of the CHSH-Bell inequality $S\leq 2$ by 3.3 standard deviations.

The experiment can be further extended to a much larger scale using photons from distant stars, e.g. the Sirius, within currently available technologies. With a single photon source of better purity[34-36], one can tolerate even feebler starlight and shorten the data accumulation time significantly (see Sec. V of Supplemental Material). Moreover, the technique reported in this work can be implemented to study astronomical objects' spectroscopy, probing into the second-order coherence information of the luminous objects (processes), e.g. single-mode line-width, decoherence properties, spectral density matrix[37]. We propose to probe the second order coherence of photons in other wavelengths from sunlight, especially the special spectra lines of elements abundant in the solar atmosphere. We expect their coherence properties may reveal dynamic sun activities, for example, dramatic changes in magnetic field, reemission of photons at dark lines, dispersion

(chirping) mechanism, etc. Besides, our techniques reported in this paper demonstrate the highest measured $g_{\text{Sun}}^{(2)}(0) = 1.94$ over all previous works on measuring $g_{\text{Sun}}^{(2)}(0)$ [38], calling for future applications in highly-sensitive starlight intensity interferometry, and large-scale ghost-imaging experiment with sunlight[39,40].

In summary, we have reported a quantum optics experiment with extremely dissimilar photons, which are emitted from a semiconductor QD and from the Sun, respectively, separated by a distance of one astronomical unit. Mixing the high-performance quantum-dot single-photon source with the thermal light, our experiment observed highly non-classical two-photon Hong-Ou-Mandel interference with ~80% raw visibility, thus providing the first and unambiguous evidence of the quantum nature of the thermal light. The two dissimilar light souces have been further used to generatehigh-fidelity entanglement and violation of Bell inequality. Our results validate the universality of the principles of quantum statistics and violation of Bell's inequality on astronomical scales.

**Figure captions**

**Figure 1 | Experimental arrangement combing quantum dot single photon source and Sun.** The InAs/GaAs QD-micropillar system lies in a cryostat at 4K, and is resonantly excited by a pulsed pico-second laser (not shown) to emit single photons in pulses which are collected by a confocal setup. The single photons then go through a polarization beam splitter (PBS), and spectrally filtering by an 1GHz-FWHM etalon. In precedence of the beam splitter (BS) for two photon interference lies a half-wave-plate (HWP) which determines whether parallel polarization (indistinguishable) case or cross polarization (distinguishable) case it is in. Sunlight is collected by a simple optical setup fixed to an equatorial mount, in precedence of which a grating is placed to select out infrared band near 893 nm. These photons from Sun are polarized at a PBS and filtered by a 1GHz-FWHM etalon. A superconducting nanowire single photon detector (SNSPD) with time-resolution of 10 ps is used to register the arrival of single photons. Detection signals are led to a time-to-digital converter (not shown) where electrically gating and coincident analysis are performed. For correlation measurement in entanglement establishment and Bell experiment, between the beam splitter and SNSPDs two polarizers are inserted (not shown).

**Figure 2 | Characteristics of the independent sources. a,** Intensity-correlation histogram of the pulsed RF from quantum dot obtained using a Hanbury Brown and Twiss-type setup. The second-order correlation $g_Q^2(0) = 0.011(1)$ is calculated by summing all coincidence counts in the zero-delay peak divided by that of its adjacent peak. The right panel is a zooming-in plot of the coincidence counts around zero time delay (green) and 26ns delay (purple) with log scale. **b,** Intensity-correlation histogram of the pulsed RF from Purcell-

enhanced QD-micropillar system in Hong-Ou-Mandel interference setup with cross polarization (red) and parallel polarization (blue). The time separation between the two single photons emitted from the single QD set to 13.2 ns (one pulse separation). The right Panel is a zooming-in plot of the coincidence counts around zero time delay. A raw visibility $V_r = (P_{\text{cross}} - P_{\text{parallel}})/P_{\text{cross}} = 96.36(4)\%$, where $P_{\text{cross (parallel)}}$ stands for the coincidence probability in the cross (parallel) polarization case, is calculated by integrating all the coincidence counts on the whole central coincidence peak. The data is normalized to the parallel polarization situation. **c,** Intensity-correlation histogram of the sunlight photons after 1ns electrically-gating obtained using a Hanbury Brown and Twiss-type setup. The second-order correlation $g_S^2(0) = 1.94(13)$ is calculated from the 40 ps-long central bin's counts of the zero-delay peak divided by that of its adjacent peaks.

**Figure 3 | HOM interference between photons from QD and the Sun. a,** Histogram of the time-delays of the coincident arrivals of the two photons on the superconducting nanowire single photon detectors. Blue disks correspond to the parallel polarization (indistinguishable) situation while red circles correspond to cross polarization situation (distinguishable). Each bin is 10 ps long. We obtain an overall coincidence count rate of 32/s for cross polarization and 24/s for the parallel polarization, and for the central 10 ps bin around zero-time delay the coincidence count rate is 0.36/s for cross polarization and 0.07/s for parallel polarization. The solid lines are fits based on the model in Supplemental Sec. II. The counts of parallel-polarization situation have been normalized to the counts of cross polarization situation for a fair comparison. Error bars indicate the standard deviation. **b,** Dependence of the raw visibility on the width of sampling delay bins centered at the

zero delay. Red horizontal dashed line corresponds to the classical limit 0.5. Error bars represent statistical errors.

**Figure 4 | Entanglement fidelity measurement of the entangled photon pairs with no common history. a, b & c,** Polarization analysis of the entangled pairs on three different bases. The plotted data are normalized raw data without subtraction of background counts or any other correction. The bars indicate the observed fractions for the counts of specific polarization observables over the total counts of all the polarization observables. The measurements are taken on three bases (HV, + −, RL) respectively to calculate the fidelity of established entanglement between photons from the Sun and QD. Error bars represent statistical errors.

**Figure 5 | Correlation functions for the CHSH inequality.** Labels on X-axis with the form of $(\phi_1, \phi_2)$ indicate the measured polarization bases of the entangled pairs of photons from the Sun and quantum dot. Error bars represent statistical errors.

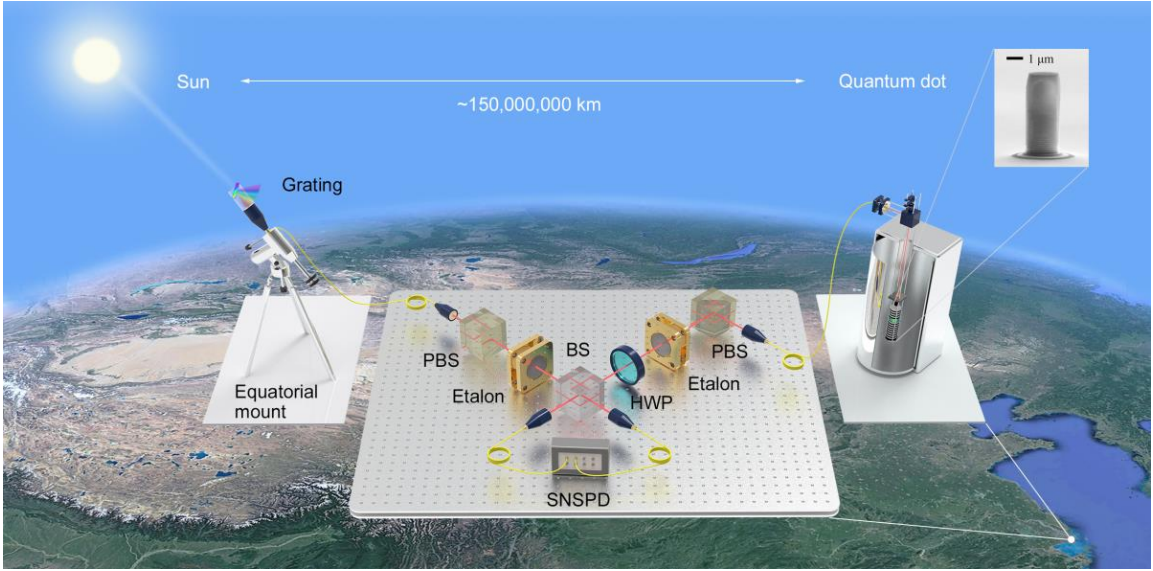

Figure 1

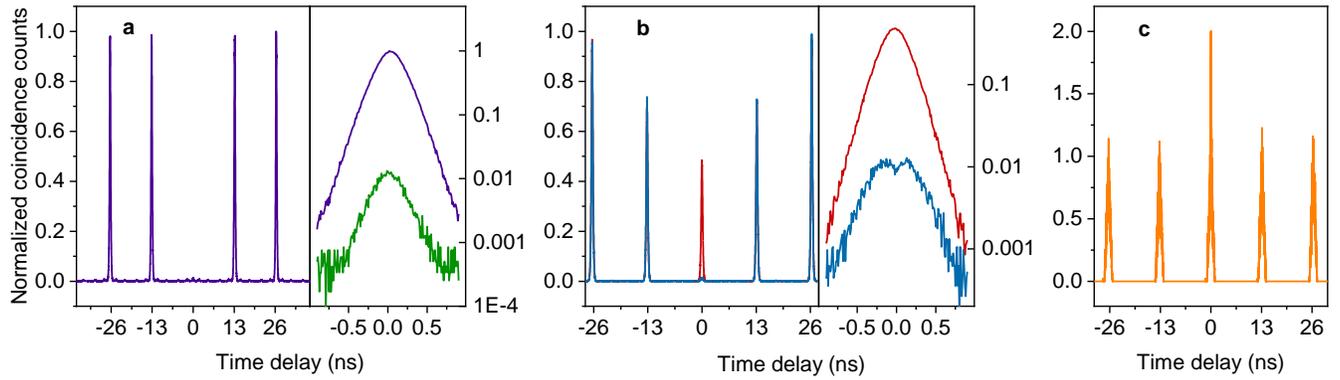

Figure 2

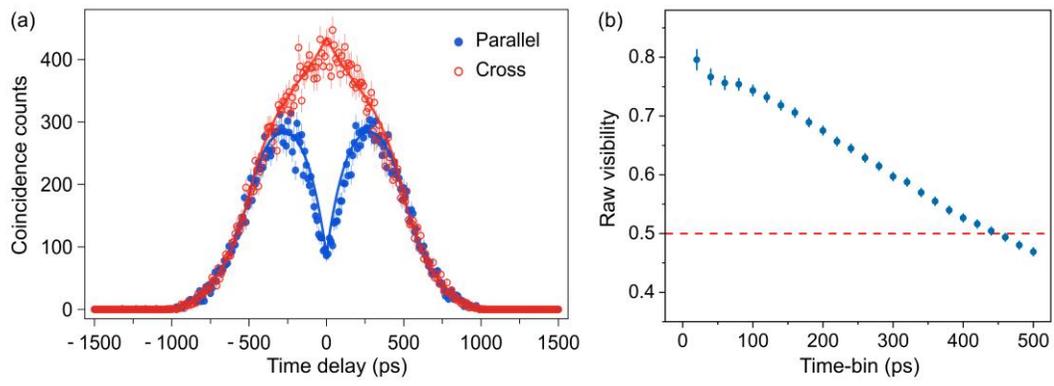

Figure 3

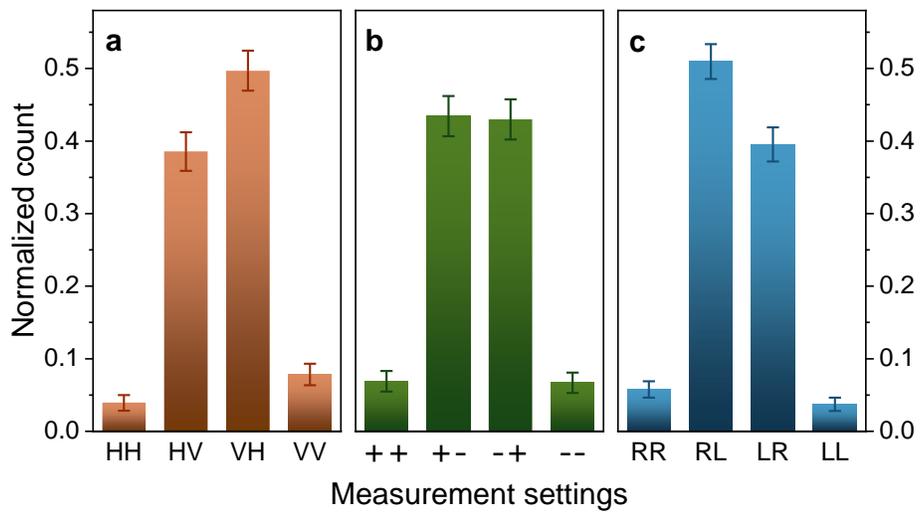

Figure 4

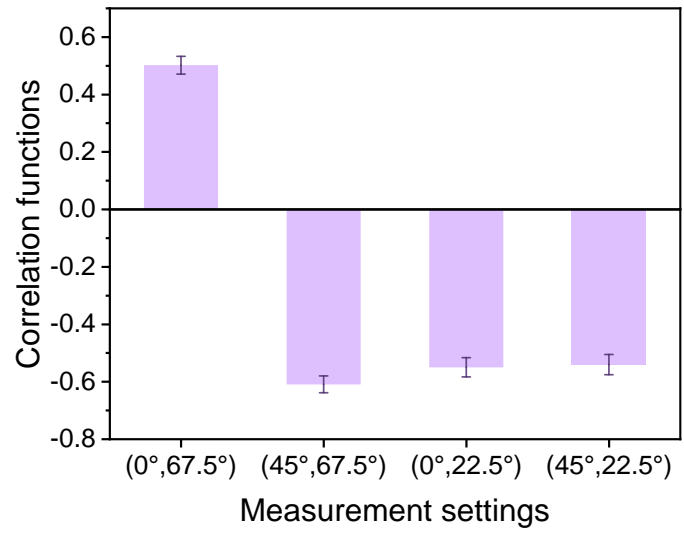

Figure 5